\documentclass[final,5p,times,twocolumn]{elsarticle}
\usepackage{graphicx}
\usepackage{bm}
\usepackage{amsmath}
\usepackage{amsfonts}
\usepackage{amssymb}
\biboptions{compress}

\journal{Physics Letters B}

\begin{document}

\begin{frontmatter}

\title{Photon structure functions at small $x$ in holographic QCD}

\author[AS]{Akira~Watanabe}

\ead{watanabe@phys.sinica.edu.tw}

\author[AS,NTHU,NCKU]{Hsiang-nan Li}

\ead{hnli@phys.sinica.edu.tw}

\address[AS]{Institute of Physics, Academia Sinica, Taipei, Taiwan 115, Republic of China}

\address[NTHU]{Department of Physics, National Tsing-Hua University, Hsinchu, Taiwan 300, Republic of China}

\address[NCKU]{Department of Physics, National Cheng-Kung University, Tainan, Taiwan 701, Republic of China}

\begin{abstract}
We investigate the photon structure functions at small Bjorken variable $x$ in the framework of the holographic QCD, assuming dominance of the Pomeron exchange. The quasi-real photon structure functions are expressed as convolution of the Brower-Polchinski-Strassler-Tan (BPST) Pomeron kernel and the known wave functions of the U(1) vector field in the five-dimensional AdS space, in which the involved parameters in the BPST kernel have been fixed in previous studies of the nucleon structure functions. The predicted photon structure functions, as confronted with data, provide a clean test of the BPST kernel. The agreement between theoretical predictions and data is demonstrated, which supports applications of holographic QCD to hadronic processes in the nonperturbative region. Our results are also consistent with those derived from the parton distribution functions of the photon proposed by Gl\"uck, Reya, and Schienbein, implying realization of the vector meson dominance in the present model setup.
\end{abstract}


\end{frontmatter}

\section{\label{sec:level1}Introduction}
A photon is a fundamental particle, instead of a nonperturbative composite like hadrons.
However, an energetic photon can fluctuate into quark-antiquark pairs, which further
radiate gluons, in a hard scattering process. Hence, investigations on the photon
structure have served as alternative tests of perturbative approaches to QCD over recent decades.
The authors in Refs.~\cite{Brodsky:1971vm,Walsh:1971xy} first discussed the feasibility
of the electron-photon deep inelastic scattering (DIS) and estimated its cross section,
whose measurements, such as those conducted at the LEP, provided valuable information of
the photon structure functions.
After Witten first evaluated the leading-order QCD corrections~\cite{Witten:1977ju},
many elaborated higher-order calculations based on perturbative QCD had been performed~\cite{LlewellynSmith:1978dc,Bardeen:1978hg,Duke:1980ij,Antoniadis:1982fv,Gluck:1983mm,
Fontannaz:1992gj,Gluck:1991ee},
and the next-to-next-to-leading-order results were available~\cite{Moch:2001im}.
The consistency between experimental data and theoretical results has strongly supported
the reliability of the perturbative calculations in QCD. The picture
of the photon structure becomes very different in the kinematic region with a small Bjorken
variable $x$. A photon can fluctuate into vector mesons, so hadronic contributions are
not negligible to electron-photon DIS as $x < 0.1$ in general. The hadronic component
dominates as $x < 0.01$, and a photon can be regarded as a hadron rather than a
pointlike object. Effective models are then appropriate for studies of
the photon structure functions in this region.

The holographic QCD based on the
AdS/CFT correspondence~\cite{Maldacena:1997re,Gubser:1998bc,Witten:1998qj} is one of the effective
models. The AdS/CFT correspondence implies that one can analyze a gauge-theory process at
strong coupling in the ordinary Minkowski space by means of the corresponding gravity theory
at weak coupling in the higher dimensional AdS space.
Since Maldacena first proposed this correspondence, its applications to QCD
processes have been attempted intensively~\cite{Kruczenski:2003be,Kruczenski:2003uq,Sakai:2004cn,
Sakai:2005yt,Son:2003et,Erlich:2005qh,DaRold:2005zs,Erdmenger:2007cm},
in which the QCD scale $\Lambda_\mathrm{QCD}$ is introduced and the conformal invariance is broken.
In particular, phenomenological applications in hadron physics to, e.g., mass spectra,
form factors, high-energy scattering processes, and so on, are quite successful.

In this letter we will investigate the photon structure functions at small $x$, adopting
the Pomeron exchange picture in the framework of the holographic QCD. A Pomeron is considered
as a color singlet gluonic state composed of multi-gluon exchanges. It has been known that
cross sections for various high-energy scattering processes are well described in this
picture~\cite{Donnachie:1992ny}. A Pomeron in QCD corresponds to a reggeized graviton in
the gravity theory in the five-dimensional AdS space. The authors in
Ref.~\cite{Brower:2006ea} evaluated the graviton exchange, and proposed the
Brower-Polchinski-Strassler-Tan (BPST) Pomeron kernel.
The elaborated analysis in Ref.~\cite{Cornalba:2007zb,Cornalba:2008sp,Brower:2007qh,Brower:2007xg,
Cornalba:2010vk,Cornalba:2009ax} has indicated that applications of this kernel to the nucleon structure
functions at small $x$ led to results well consistent with
data~\cite{Brower:2010wf,Watanabe:2012uc,Watanabe:2013spa}.

Following the same formalism, we will express the photon structure functions at small $x$ as
convolution of the BPST kernel with the wave functions of the
incident and target photons in the 5th dimension. Most parameters in the BPST kernel have
been fixed in the previous studies of the nucleon structure
functions~\cite{Watanabe:2012uc,Watanabe:2013spa}. The known five-dimensional
U(1) vector field will be adopted for the photon wave functions, which couple to
leptons at the ultraviolet (UV) boundary. That is, we do not need approximations or models
to describe the incident and target particles here, in contrast to the case of the nucleon
DIS. Therefore, our study of the photon structure functions involves only a single
free parameter, which appears as an overall coefficient for the cross section. The model dependence
is then reduced, and the predicted photon structure functions, as confronted with data, serve
as a clean check of the validity of the BPST kernel.
It will be demonstrated that our results for the photon structure function $F_2^\gamma (x,Q^2)$
are in agreement with the LEP data, though available data at small $x$ are limited.
In addition, our results are also consistent with those obtained from the parton distribution
functions (PDFs) of the photon proposed by Gl\"uck, Reya, and Schienbein~\cite{Gluck:1999ub}.
Because they included the hadronic component in the photon PDFs, which dominates
at small $x$, this consistency may imply realization of the vector meson dominance
in the present setup. Our predictions for the photon structure functions at very low $x$
can be tested at future linear colliders.

\section{\label{sec:level2}Kinematics and model setup}
\begin{figure}[bt!]
\begin{center}
\includegraphics[width=77mm]{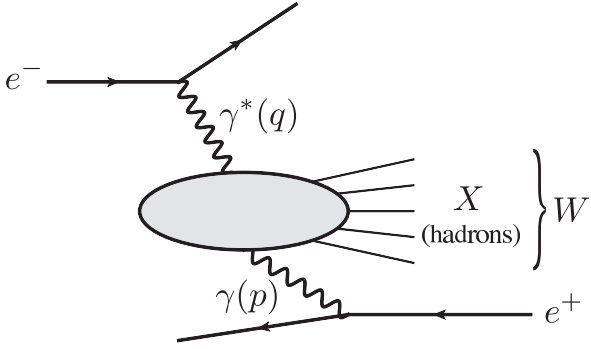}
\caption{
Deeply inelastic electron-photon scattering process, where
the bubble represents a parton density of the target photon.
}
\label{fig:electron-photon_DIS}
\end{center}
\end{figure}

The electron-photon DIS $e\gamma  \to eX$ is schematically shown in Fig.~\ref{fig:electron-photon_DIS},
where $q$ and $p$ are the four-momenta of the probe and target photons, respectively,
and $W$ represents the invariant mass of the hadronic final state.
The associated differential cross section is expressed in terms of the two
structure functions $F_2^\gamma (x,Q^2)$ and $F_L^\gamma (x,Q^2)$ as
\begin{equation}
\frac{{{d^2}{\sigma _{e\gamma  \to eX}}}}{{dxd{Q^2}}} =
\frac{{2\pi {\alpha ^2}}}{{x{Q^4}}}\left\{ \left[ {1 + {{\left( {1 - y} \right)}^2}}
\right]F_2^\gamma \left( {x,{Q^2}} \right) - {y^2}F_L^\gamma \left( {x,{Q^2}} \right) \right\},\label{DIS}
\end{equation}
where $\alpha$ is the fine structure constant, $y$ is the inelasticity, and the
Bjorken scaling variable $x$ is defined as
\begin{equation}
x = \frac{Q^2}{Q^2 + W^2 + P^2},
\end{equation}
with $Q^2 = -q^2$ and $P^2 = -p^2$. We consider the small $x$ region with the
target being a quasi-real photon, namely, the kinematic region of $W^2 \gg Q^2 \gg P^2$,
in which the Bjorken variable can be approximated by $x \approx Q^2/W^2$.
The vector meson dominance model~\cite{Sakurai:1960ju} appropriate in this
region implies that the target photon may behave like a vector
meson rather than a point particle, and suggests the application of the
Pomeron exchange picture to this process.

The photon structure functions in Eq.~(\ref{DIS}) are given in the five-dimensional
AdS space by~\cite{Brower:2007qh,Brower:2007xg},
\begin{align}
&F^{\gamma}_{i} (x,Q^2) = \frac{\alpha g_0^2 \rho^{3/2}Q^2}{32 \pi ^{5/2}}
\int dzdz' P_{13}^{(i)} (z,Q^2) P_{24}(z',P^2) \nonumber \\
&\hspace{43mm} \times (zz') \mbox{Im} [\chi_{c} (W^2,z,z')], \label{eq:Fic}
\end{align}
with $i = 2,L$, where the adjustable
parameters $g_0^2$ and $\rho$ control the magnitude and the energy
dependence of the structure functions, respectively. The imaginary part of the
BPST Pomeron kernel~\cite{Brower:2006ea,Brower:2007xg}
\begin{eqnarray}
\mbox{Im} [\chi_{c} (W^2,z,z') ] = e^{(1-\rho)\tau}
e^ {-[({\log ^2 z/z'})/{\rho \tau}]} / \sqrt{\tau}, \label{eq:kernel_c}
\end{eqnarray}
with the conformal invariant
\begin{eqnarray}
\tau = \log (\rho z z' W^2 /2),\label{eq:tau}
\end{eqnarray}
arises from the single-Pomeron/graviton exchange.

The overlap functions $P_{13}(z)$ and $P_{24}(z')$, describing the density distributions
of the incident and target particles in the AdS space, respectively,
obey the normalization condition $\int dz P_{ij}(z) = 1$ in the on-shell case.
They depend on the particle virtualities $Q^2$ and $P^2$ in the present study
with off-shell photons. $P_{13}^{(L)}(z,Q^2)$ denotes the wave
function of the longitudinally polarized photon, and $P_{24}(z',P^2)$ takes the
form the same as $P_{13}^{(2)}(z,Q^2)$.
The massless 5D U(1) vector field, satisfying
the Maxwell equation in the five-dimensional AdS background spacetime, can be
identified as the physical photon at the UV boundary. We adopt the wave function of
this field for the overlap functions, which are then written as~\cite{Polchinski:2002jw}
\begin{align}
&P_{13}^{(2)} (z,Q^2) = Q^2 z \left[ K_0^2(Qz) + K_1^2(Qz) \right], \label{eq:P132} \\
&P_{13}^{(L)} (z,Q^2) = Q^2 z K_0^2(Qz). \label{eq:P13L}
\end{align}
Both $K_0 (Qz)$ and $K_1 (Qz)$, the modified Bessel functions of
the second kind, diverge at the origin and vanish at large $Qz$,
consistent with the fact that a photon is a non-normalizable mode, and can
penetrate into the larger $z$, i.e., infrared (IR) region at low $Q$. Namely,
Eq.~(\ref{eq:P132}) exhibits the feature that a quasi-real
photon propagates into the bulk of the AdS space, while keeping a substantial
component in the UV region even at small $x$.

In the considered region with $10^{-5} \leq x \leq 10^{-2}$ and
1~GeV$^2$ $\leq Q^2 \leq 10$~GeV$^2$, nonperturbative hadronic contribution to the
structure functions may become important.
We employ the modified kernel~\cite{Brower:2006ea,Brower:2010wf}
\begin{align}
&\mbox{Im} [\chi_{mod} (W^2,z,z')] \equiv \mbox{Im} [\chi_c (W^2,z,z') ] \nonumber \\
&\hspace{30mm} + \mathcal{F} (z,z',\tau) \mbox{Im} [\chi_c(W^2,z,z_0^2/z') ],\label{eq:kernel_mod} \\
&\mathcal{F} (z,z',\tau) = 1 - 2 \sqrt{\rho \pi \tau} e^{\eta^2} \mbox{erfc}( \eta ), \\
&\eta = \left( -\log \frac{zz'}{z_0^2} + \rho \tau \right) / {\sqrt{\rho \tau}},
\end{align}
where the first term on the right-hand side of Eq.~\eqref{eq:kernel_mod} is the conformal
kernel, and the second term represents the
additional confinement effect. The adjustable parameter $z_0 \sim 1/\Lambda_{\rm QCD}$,
being of order of the QCD scale, governs the strength of the confinement effect.
The factor $\mathcal{F}$, running from $1$ to $-1$ for fixed $z$ and $z'$, as $\tau$ increases
from zero to infinity, suppresses the structure functions at large $W^2$. That is, the
confinement effect intensifies at small $x$ for fixed $Q^2$.
Strictly speaking, there must be modification of the background metric as
the confinement effect is not negligible, which may lead to a more complicated
kernel with more parameters. However, it has been also observed~\cite{Watanabe:2013spa} that the
modified kernel in Eq.~(\ref{eq:kernel_mod}) well reproduces the data for the nucleon
structure functions at small $x$, which are typical nonperturbative
quantities.

\section{\label{sec:level3}Numerical results}
We take the parameter sets for the BPST kernels obtained in
Ref.~\cite{Watanabe:2013spa} with $z_0 = 4.25$~GeV$^{-1}$ or $z_0 = 6$~GeV$^{-1}$ 
in the modified kernel case,
which has been demonstrated to well reproduce the data of the nucleon structure
functions at small $x$. Since a photon is a
non-normalizable mode, the overall coefficient $g_0^2$ in Eq.~\eqref{eq:Fic} is an
unknown, and has to be tuned to fit the absolute magnitude of data. The available
data of the real photon structure functions at small $x$ are still quite
limited. We consider 9 data points from the OPAL collaboration at
LEP~\cite{Abbiendi:2000cw} with $x \leq 0.025$, and determine $g_0^2 = 17.51$ and
$49.01$ for the conformal and modified kernels, respectively. The invariant mass squared
$P^2$ of the target photon should be small enough, but cannot be exactly zero in practice.
Events in the region with $P^2 \sim \mathcal{O} (0.01)$~GeV$^2$
have been measured (for details, see review papers~\cite{Nisius:1999cv,Krawczyk:2000mf}
and references therein), so $P^2 = 0.01$~GeV$^2$ is assumed here~\cite{Gluck:1999ub}.

\begin{figure}[bt!]
\begin{center}
\includegraphics[width=85mm]{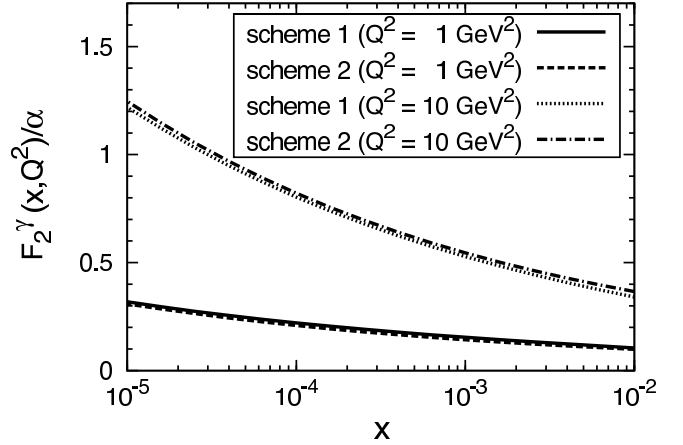}
\caption{
$F_2^\gamma (x,Q^2)$ as a function of $x$ for $Q^2 = 1, 10$~GeV$^2$ derived
from schemes 1 and 2 for the treatment of the conformal invariant $\tau$ with the modified kernel.
}
\label{fig:scheme1_vs_sheme2}
\end{center}
\end{figure}

We explain the treatment of the conformal invariant $\tau$, which appears
in the kernels in Eqs.~(\ref{eq:kernel_c}) and
(\ref{eq:kernel_mod}). Straightforward integrations collect the contribution from
the region of small $z$ and $z'$, where the photon distributions concentrate,
yielding a negative $\tau$, i.e., an imaginary piece in the kernels.
A simple resolution of this difficulty is to restrict $z$ and $z'$ in the region,
where $\tau$ remains positive. We refer to this prescription as scheme 1.
Another way is to set $z$ and $z'$ in the definition of $\tau$ to their average
values defined by
\begin{equation}
{\bar{z}} \equiv \frac{{\int {dz} {z^2}{P_{13}^{(i)}}
\left( {z,{Q^2}} \right)}}{{\int {dzz{P_{13}^{(i)}}
\left( {z,{Q^2}} \right)} }}, \ \ \ \ \ {\bar{z}'}
\equiv \frac{{\int {dz'} {{z'}^2}{P_{24}}
\left( {z',{P^2}} \right)}}{{\int {dz'z'{P_{24}}\left( {z',{P^2}} \right)} }},
\end{equation}
respectively, and regarding the product $zz'$ in Eq.~\eqref{eq:tau} as the characteristic
scale $\bar z\bar z'$ of the kernels. We have verified that this prescription, referred
to as scheme 2, works for explaining the data of the nucleon structure functions with the
model setup in Ref.~\cite{Watanabe:2013spa}.
As indicated in Fig.~\ref{fig:scheme1_vs_sheme2}, the results for the photon structure function
$F_2^\gamma (x,Q^2)$ from these two schemes differ slightly (about few percents) at least in the
currently considered kinematic region.
We will adopt scheme 2 for the numerical analysis below.

\begin{figure*}[bt!]
\begin{center}
\includegraphics[width=145mm]{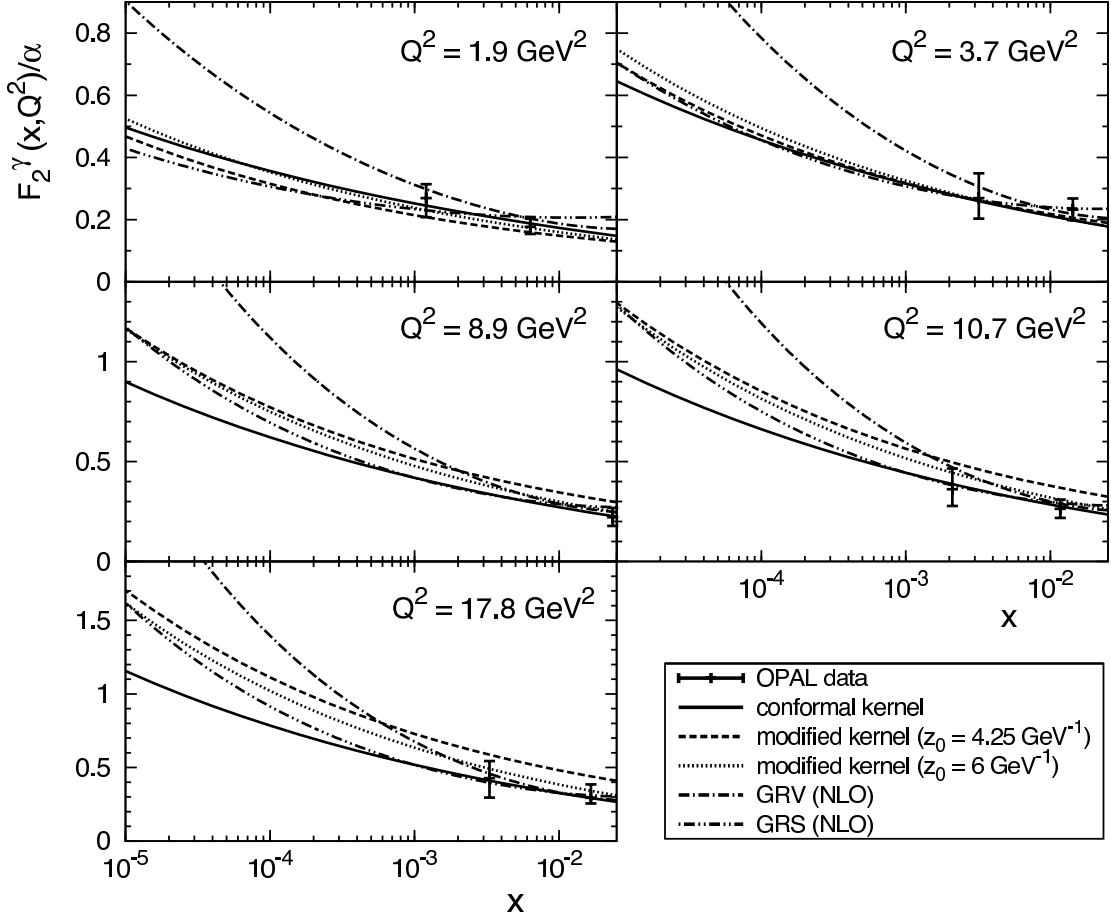}
\caption{
$F_2^\gamma (x,Q^2)$ as a function of $x$ for various $Q^2$,
and its comparison with the data from the OPAL Collaboration~\cite{Abbiendi:2000cw}.
In each panel, the solid, dashed, and dotted curves are from our calculations.
The dash-dotted and dashed double-dotted curves represent the GRV~\cite{Gluck:1991jc}
and GRS~\cite{Gluck:1999ub} predictions, respectively.
}
\label{fig:F2_photon}
\end{center}
\end{figure*}
We display in Fig.~\ref{fig:F2_photon} the predicted $x$ and $Q^2$ dependencies
of the structure function $F_2^\gamma$ for the quasi-real
photon from Eq.~\eqref{eq:Fic} with the conformal and modified Pomeron
kernels, and those derived from the well-known PDF sets,
GRV~\cite{Gluck:1991jc} and GRS~\cite{Gluck:1999ub}, at next-to-leading-order accuracy.
Both curves from GRV and GRS include the charm-quark contribution in the region with
$W \geq 2 m_c $. Note that the hadronic contributions to the GRV and GRS PDFs were
parameterized via the pion PDFs based on the assumed similarity between the vector meson
and the pion, and that the validity of the parameterizations in
Refs.~\cite{Gluck:1991jc,Gluck:1999ub} is limited to the range $Q^2 > 0.5$~GeV$^2$
and $x > 10^{-5}$. It is seen in Fig.~\ref{fig:F2_photon} that all the above curves
basically match the data available for $x>10^{-3}$, and then ascend with
different slopes as $x$ decreases, a feature attributed to the Pomeron exchange.
Future data will further discriminate the different predictions at small $x< 10^{-3}$.
To be precise, the results from the modified kernel with $z_0 = 6$~GeV$^{-1}$ and
from the conformal kernel fit the data equally well based on $\chi^2$ per degree of freedom, 
and slightly better than those from the modified kernel with $z_0 = 4.25$~GeV$^{-1}$.
If we focus on the region with low $Q^2 < 10$~GeV$^2$ and small $x < 0.01$, and
tune the only parameter $g_0^2$ again, the modified kernel with $z_0 = 6$~GeV$^{-1}$ gives 
the best fit to the data.

The results from the conformal kernel (GRV PDFs) show
the weakest (strongest) $x$ dependence, and those from the GRS PDFs show the moderate
$x$ dependence.  The obvious difference between the GRV and GRS predictions in the small
$x$ region, and their comparison with the OPAL data have been discussed in Ref.~\cite{Gluck:2001nx}.
The curves from the modified kernel exhibit the similar $x$ dependence
and the stronger $Q^2$ dependence compared to those from the conformal kernel.
The former are closer to the GRS curves compared to the latter, and coincide with
the GRS curves as indicated in the first two panels of Fig.~\ref{fig:F2_photon}.
It is straightforward to find that the hadronic component
in the GRS photon PDFs dominates at small $x$, about 90\% (80\%) of the total contribution to
$F_2^\gamma$ for $Q^2=1$ (10) GeV$^2$. This consistency may imply realization of the vector
meson dominance in the present model setup.

It is found that the difference between the curves from the modified kernel
and from the conformal kernel becomes larger, as $Q^2$ increases and as $x$ decreases.
Note that the two kernels in Eqs.~(\ref{eq:kernel_c}) and (\ref{eq:kernel_mod})
have been treated as independent models, and their adjustable parameters
have been fixed separately~\cite{Watanabe:2013spa}. It has been known that the different
parameter sets are needed to well describe the data of the nucleon structure function
$F_2^p$ with the two kernels in the $Q^2 < 10$~GeV$^2$ region.
Therefore, the growing difference does not imply that the confinement
effect from the second term in Eq.~\eqref{eq:kernel_mod} increases with $Q^2$.
It does not imply either that this term is positive in the small $x$ region.
Choosing a common parameter set for both the conformal and modified kernels,
we indeed see that the relative difference between the resultant photon structure
functions reduces, and the latter become lower than the former as $x$ decreases.
That is, the confinement effect is weakened at high $Q^2$, and the second term
in Eq.~\eqref{eq:kernel_mod} remains negative at small $x$. Nevertheless,
the resultant photon structure functions approach to each other at high $Q^2$ more
slowly than in Ref.~\cite{Brower:2010wf} due to the different treatments of density
distributions of the involved particles\footnote{The kernel ratio
$r\equiv \mbox{Im} [\chi_{mod}]/\mbox{Im} [\chi_{c}]$, which
reflects the detailed difference between the modified and conformal kernels,
has been investigated, and its behavior has been shown in Fig.~1 of
Ref.~\cite{Watanabe:2013spa}.}.

The agreement of the results from the modified kernel with the data can be further improved
by considering the number of active quark flavors $f$ involved in the analysis.
The target nucleon overlap function is localized in the IR region around
$z' \sim 4$~GeV$^{-1}$~\cite{Watanabe:2013spa}, so that the heavy quark contribution is
suppressed, and we have $f \sim 4$. The target photon overlap function $P_{24} (z',P^2)$
spreads broadly from the UV region to the IR region in the $z'$ space, implying the
participation of all flavors, i.e., $f =  6$. Because
the QCD scale $\Lambda_{\rm QCD}\sim 1/z_0$ decreases with $f$,
the parameter $z_0$ in the present study may be larger than that fixed in
Ref.~\cite{Watanabe:2013spa}. We then calculate the structure function $F_2^\gamma$,
assuming $z_0 = 6$~GeV$^{-1}$ and $g_0^2 = 21.60$, and display the results
in Fig.~\ref{fig:F2_photon} as well. It is observed that the corresponding dotted curves,
becoming higher than the dashed ones as $Q^2=1.9$ GeV$^2$, and lower as
$Q^2>8.9$ GeV$^2$, agree completely with the OPAL data within errors.

\begin{figure}[bt!]
\begin{center}
\includegraphics[width=85mm]{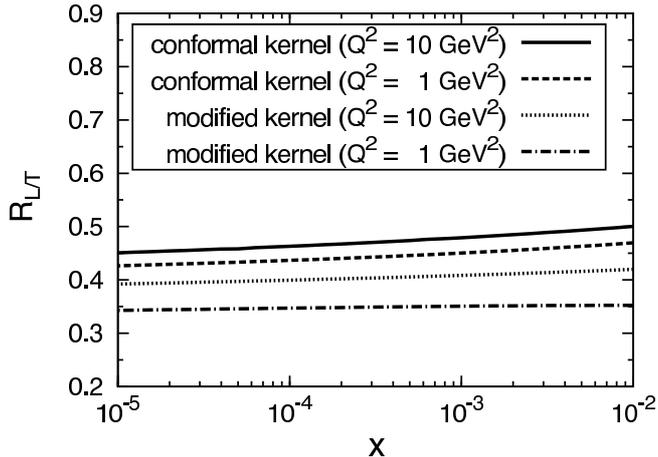}
\caption{
Ratio $R_{L/T} = F_L^\gamma (x,Q^2) / F_T^\gamma (x,Q^2)$ as a function of $x$ for $Q^2 = 1, 10$~GeV$^2$.
The parameter value $z_0 = 4.25$~GeV$^{-1}$ is adopted to obtain the curves from
the modified kernel.
}
\label{fig:RLTvsx}
\end{center}
\end{figure}

Next we present the $x$ dependence of the ratio of the longitudinal structure function
over the transverse structure function, defined as
$R_{L/T} = F_L^\gamma (x,Q^2) / F_T^\gamma (x,Q^2)$ with $ F_T^\gamma = F_2^\gamma - F_L^\gamma$,
in Fig~\ref{fig:RLTvsx}.
It is found that $R_{L/T}$ increases with $x$ and $Q^2$ in both cases of the conformal and
modified kernels.
$R_{L/T}$ seems to be almost constant for the modified kernel with $Q^2=1$ GeV$^2$.
These results are qualitatively consistent with those of the nucleon structure functions~\cite{Watanabe:2013spa}.
It is easy to understand $R_{L/T} < 1$, since the contribution from the longitudinal polarization
of the probe photon is suppressed at low $Q^2$, and the increase of $R_{L/T}$ with $Q^2$, since
this contribution increases with $Q^2$. However, $R_{L/T}$ around 0.4 also indicates that the
longitudinal structure function is not negligible in the considered kinematical region.

Finally, we compare the $x$ dependence of the photon structure function $F_2^\gamma$ with
that of the nucleon structure function $F_2^N$ by investigating
their ratio $R_{\gamma / N} = F_2^\gamma (x,Q^2) / [\alpha F_2^N (x,Q^2)]$.
\begin{figure}[bt!]
\begin{center}
\includegraphics[width=85mm]{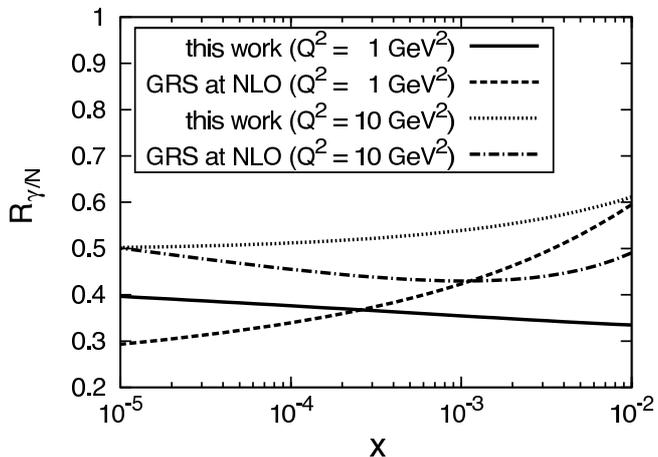}
\caption{
Ratio $R_{\gamma / N} = F_2^\gamma (x,Q^2) / [\alpha F_2^N (x,Q^2)]$ as a function of $x$ for $Q^2 = 1$
and 10~GeV$^2$, and its comparison with the GRS predictions~\cite{Gluck:1999ub}.
$F_2^\gamma (x,Q^2)$ is obtained by employing the modified kernel with $z_0 = 4.25$~GeV$^{-1}$.
$F_2^N (x,Q^2)$ is from Ref.~\cite{Watanabe:2013spa}, and common to all the curves.
}
\label{fig:F2ratio}
\end{center}
\end{figure}
For $F_2^N (x,Q^2)$, we adopt the results in Ref.~\cite{Watanabe:2013spa}, which match
the HERA data~\cite{Aaron:2009aa} well.
As shown in Fig.~\ref{fig:F2ratio}, both curves from our predictions are nearly linear
with small slopes. The similar $x$ dependencies of the two structure functions is again
attributed to the universal Pomeron kernel in our framework.
We also plot the GRS predictions for the ratio $R_{\gamma / N}$ in Fig.~\ref{fig:F2ratio}.
Our and GRS predictions are closer to each other for $Q^2=10$ GeV$^2$, because
both agree with the data at this scale as indicated in Fig~\ref{fig:F2_photon}.
Our and GRS predictions differ significantly for $Q^2=1$ GeV$^2$, especially at $x\sim 10^{-2}$,
because the former exhibits a substantial $x$ dependence, while the latter becomes almost
independent of $x$ as $x > 10^{-3}$.

\section{\label{sec:level4}Summary}
In this letter we have investigated the structure functions $F_2^\gamma$ and
$F_L^\gamma$ of the quasi-real photon in the region of small Bjorken variable
$x$ in a holographic QCD model. They were calculated by
convoluting the BPST Pomeron kernel with the wave functions of
the U(1) vector field in the five-dimensional AdS space.
Two of three adjustable parameters of the kernel have been fixed in the previous
studies on the nucleon DIS, so only a single overall parameter was left free, and
tuned to fit the absolute magnitude of the experimental data.
The predicted dependencies of $F_2^\gamma (x, Q^2)$ on $x$ and $Q^2$ are in agreement
with the OPAL data, implying the predictive power of the present model setup,
and that the Pomeron exchange picture can describe general
DIS processes at small $x$ reasonably.

We then compared our results with the structure functions derived
from the GRV and GRS PDF sets of the photon, and the consistency with the latter
was also observed. Since the hadronic component in the photon PDFs dominates
at small $x$, this consistency may be regarded as realization of the vector meson dominance.
It is known that the vector meson dominance model well accommodates experimental data
involving quasi-real photons, whose mechanism is, however, not completely understood due to
its nonperturbative origin. The wave functions for a quasi-real photon proposed in
Eq.~(\ref{eq:P132}), with broad distribution in the $z$ space,
can overlap with the $\rho$ meson wave function significantly. It is likely that the
quasi-real photon wave function and the $\rho$ meson wave function, as convoluted
with the universal Pomeron kernel, yield similar cross sections. Then
holographic models might have provided an alternative
viewpoint to realize the vector meson dominance mechanism.
Future linear colliders, such as the planned International Linear Collider,
will help fully understand the nature of a photon, which is one of the most fundamental
issues in high energy physics.

\section*{Acknowledgements}
We acknowledge Yoshio Kitadono, Tsuneo Uematsu, and Takahiro Sawada for valuable comments and discussions.
This work was supported in part
by the Ministry of Science and Technology of R.O.C. under Grant Nos.
NSC-101-2112-M-001-006-MY3 and MOST-103-2811-M-001-030.


\begin{thebibliography}{99}
\expandafter\ifx\csname url\endcsname\relax
  \def\url#1{\texttt{#1}}\fi
\expandafter\ifx\csname urlprefix\endcsname\relax\def\urlprefix{URL }\fi
\expandafter\ifx\csname href\endcsname\relax
  \def\href#1#2{#2} \def\path#1{#1}\fi

\bibitem{Brodsky:1971vm}
S.~J. Brodsky, T.~Kinoshita, H.~Terazawa,
Phys.Rev.Lett. 27 (1971) 280.

\bibitem{Walsh:1971xy}
T.~Walsh,
Phys.Lett. B36 (1971) 121.

\bibitem{Witten:1977ju}
E.~Witten,
Nucl.Phys. B120 (1977) 189.

\bibitem{LlewellynSmith:1978dc}
C.~Llewellyn~Smith,
Phys.Lett. B79 (1978) 83.

\bibitem{Bardeen:1978hg}
W.~A. Bardeen, A.~J. Buras,
Phys.Rev. D20 (1979) 166.

\bibitem{Duke:1980ij}
D.~Duke, J.~Owens,
Phys.Rev. D22 (1980) 2280.

\bibitem{Antoniadis:1982fv}
I.~Antoniadis, G.~Grunberg,
Nucl.Phys. B213 (1983) 445.

\bibitem{Gluck:1983mm}
M.~Gluck, E.~Reya,
Phys.Rev. D28 (1983) 2749.

\bibitem{Fontannaz:1992gj}
M.~Fontannaz, E.~Pilon,
Phys.Rev. D45 (1992) 382.

\bibitem{Gluck:1991ee}
M.~Gluck, E.~Reya, A.~Vogt,
Phys.Rev. D45 (1992) 3986.

\bibitem{Moch:2001im}
S.~Moch, J.~Vermaseren, A.~Vogt,
Nucl.Phys. B621 (2002) 413.

\bibitem{Maldacena:1997re}
J.~M. Maldacena,
Adv.Theor.Math.Phys. 2 (1998) 231.

\bibitem{Gubser:1998bc}
S.~Gubser, I.~R. Klebanov, A.~M. Polyakov,
Phys.Lett. B428 (1998) 105.

\bibitem{Witten:1998qj}
E.~Witten,
Adv.Theor.Math.Phys. 2 (1998) 253.

\bibitem{Kruczenski:2003be}
M.~Kruczenski, D.~Mateos, R.~C. Myers, D.~J. Winters,
JHEP 0307 (2003) 049.

\bibitem{Kruczenski:2003uq}
M.~Kruczenski, D.~Mateos, R.~C. Myers, D.~J. Winters,
JHEP 0405 (2004) 041.

\bibitem{Sakai:2004cn}
T.~Sakai, S.~Sugimoto,
Prog.Theor.Phys. 113 (2005) 843.

\bibitem{Sakai:2005yt}
T.~Sakai, S.~Sugimoto,
Prog.Theor.Phys. 114 (2005) 1083.

\bibitem{Son:2003et}
D.~T.~Son, M.~A.~Stephanov,
Phys.Rev. D69 (2004) 065020.

\bibitem{Erlich:2005qh}
J.~Erlich, E.~Katz, D.~T. Son, M.~A. Stephanov,
Phys.Rev.Lett. 95 (2005) 261602.

\bibitem{DaRold:2005zs}
L.~Da~Rold, A.~Pomarol,
Nucl.Phys. B721 (2005) 79.

\bibitem{Erdmenger:2007cm}
J.~Erdmenger, N.~Evans, I.~Kirsch, E.~Threlfall,
Eur.Phys.J. A35 (2008) 81.

\bibitem{Donnachie:1992ny}
A.~Donnachie, P.~Landshoff,
Phys.Lett. B296 (1992) 227.

\bibitem{Brower:2006ea}
R.~C. Brower, J.~Polchinski, M.~J. Strassler, C.-I. Tan,
JHEP 0712 (2007) 005.

\bibitem{Cornalba:2007zb}
L.~Cornalba, M.~S. Costa, J.~Penedones,
JHEP 0709 (2007) 037.

\bibitem{Cornalba:2008sp}
L.~Cornalba, M.~S. Costa,
Phys.Rev. D78 (2008) 096010.

\bibitem{Brower:2007qh}
R.~C. Brower, M.~J. Strassler, C.-I. Tan,
JHEP 0903 (2009) 050.

\bibitem{Brower:2007xg}
R.~C. Brower, M.~J. Strassler, C.-I. Tan,
JHEP 0903 (2009) 092.

\bibitem{Cornalba:2010vk}
L.~Cornalba, M.~S. Costa, J.~Penedones,
Phys.Rev.Lett. 105 (2010) 072003.

\bibitem{Cornalba:2009ax}
L.~Cornalba, M.~S. Costa, J.~Penedones,
JHEP 1003 (2010) 133.

\bibitem{Brower:2010wf}
R.~C. Brower, M.~Djuric, I.~Sarcevic, C.-I. Tan,
JHEP 1011 (2010) 051.

\bibitem{Watanabe:2012uc}
A.~Watanabe, K.~Suzuki,
Phys.Rev. D86 (2012) 035011.

\bibitem{Watanabe:2013spa}
A.~Watanabe, K.~Suzuki,
Phys.Rev. D89 (2014) 115015.

\bibitem{Gluck:1999ub}
M.~Gluck, E.~Reya, I.~Schienbein,
Phys.Rev. D60 (1999) 054019.

\bibitem{Sakurai:1960ju}
J.~Sakurai,
Annals Phys. 11 (1960) 1.

\bibitem{Polchinski:2002jw}
J.~Polchinski, M.~J. Strassler,
JHEP 0305 (2003) 012.

\bibitem{Abbiendi:2000cw}
G.~Abbiendi, et~al.,
Eur.Phys.J. C18 (2000) 15.

\bibitem{Nisius:1999cv}
R.~Nisius,
Phys.Rept. 332 (2000) 165.

\bibitem{Krawczyk:2000mf}
M.~Krawczyk, A.~Zembrzuski, M.~Staszel,
Phys.Rept. 345 (2001) 265.

\bibitem{Gluck:1991jc}
M.~Gluck, E.~Reya, A.~Vogt,
Phys.Rev. D46 (1992) 1973.

\bibitem{Gluck:2001nx}
M.~Gluck, E.~Reya, I.~Schienbein,
Phys.Rev. D64 (2001) 017501.

\bibitem{Aaron:2009aa}
F.~Aaron, et~al.,
JHEP 1001 (2010) 109.

\end{thebibliography}
\end{document}